# Mapping evolving population geography in China:
## Spatial redistribution, regional disparity, and urban sprawl

Lei Dong[1*], Rui Du[2], and Yu Liu[1*]

1. Institute of Remote Sensing and Geographical Information Systems, School of Earth and Space Sciences, Peking University, Beijing, China
2. Department of Economics, Spears School of Business, Oklahoma State University, Stillwater, U.S.

Corresponding author. Email: leidong@pku.edu.cn (L.D.)  liuyu@urban.pku.edu.cn (Y.L.)

**Abstract:** China's demographic changes have important global economic and geopolitical implications. Yet, our understanding of such transitions at the micro-spatial scale remains limited due to spatial inconsistency of the census data caused by administrative boundary adjustments. To fill this gap, we manually collected and built a population census panel from 2010 to 2020 at both the county and prefectural-city levels. We show that the massive internal migration drives China's increasing population concentration and regional disparity, resulting in severe population aging in shrinking cities and increasing gender imbalance in growing cities.

**Main Text:** As the world's most populous country, China's demographic transition is not only a domestic issue but also a highly charged geopolitical and economic one that will have profound global impacts. Over the past decade, China's population has peaked and is expected to reach a critical demographic turning point in the foreseeable future. Population transitions, such as sluggish population growth, falling birth rates, rapid aging, massive migration and urbanization, and the disappearing "demographic dividend", will impose far-reaching socioeconomic consequences. These are arguably some of the most pressing issues policymakers, researchers, and businesses face today (*1, 2*). For instance, China's net population growth was only 0.48 million last year, which is far lower than expected and signals an early onset of China's population peak (*3*). However, despite a great deal of evidence on the overall evolution of demographics (*4-7*), demographic transitions at the micro-spatial scale are understudied. The micro-spatial patterns of demographic changes have significant implications for population and regional policies required for sustainable long-run economic growth.

Most countries rely on census data to seek a comprehensive probe into population dynamics. However, the records between census years at the city/county level (*8*) in China are not comparable as the country has made substantial changes to administrative levels and boundaries over the past decade. Furthermore, these adjustments were not well documented in the released census datasets. To overcome such data challenges, we manually collected and digitized 2,317 gazettes (~10,000 documents in total, fig. S1) from local official sources for the 2020 Census and then merged them with 2010 Census data. To consider intercensal changes in county boundaries, we matched county-level data with 770 released administrative changes, e.g., changes in the status of the administrative level, city/county consolidation or disintegration, land reconfigurations, etc. Data collection and processing are detailed in the supplementary materials (SM). Finally, we built a population panel from 2010 to 2020 at both the county and prefectural-city levels, which covers 2,666 counties in 356 prefectural cities (*8*). The dataset has been made publicly available to facilitate future research.

Using self-built county-level census data, we quantify the major changes in China's population geography. We find that the massive internal migration is the key determinant of China's increasing population concentration and regional disparity, which are closely related to population aging and gender imbalance.

**MAPPING POPULATION GROWTH AND DECLINE**

Figure 1 shows three important geographical patterns of the population distribution, most likely due to spatially heterogeneous migration:

**Population distribution has become more concentrated, where we see fewer growing counties and more shrinking ones.** Of the 2,666 counties, 45.8% experienced an increase

in population and 54.2% experienced a decline. Less than 33.5% of the counties grew faster than the national average growth rate of 5%, with nearly two-thirds of counties having a population growth rate below the national average. We also observe considerable regional heterogeneity in population growth. The coastal provinces have the largest population growth, such as Guangdong [+21.7millon (m) persons], Zhejiang (+10.1m), and Jiangsu (+6.1m), whereas northeastern provinces witnessed the most severe population outflow. The share of counties with declining populations amounted to 85.2% in Heilongjiang (-6.5m), 88.7% in Jilin (-3.4m), and 76.3% in Liaoning (-1.2m). In agreement with this concentration trend, the Gini coefficient of county-level population distribution increased from 0.403 in 2010 to 0.437 in 2020 (fig. S2).

**Population migration pattern reveals a hierarchical structure.** The most salient trend is that the population flowed from neighboring counties to provincial capitals within provinces, with several mega-cities (cities with more than 10m population) drawing the largest population inflow at the national scale (e.g., Shenzhen, the city with the largest population increase from 2010 to 2020, grew by a total of 7.1 million). Of the 32 cities with a net population inflow of over a million, twenty are provincial capitals or municipalities directly administered by the central government. The rest are generally the most economically developed cities in each province, such as Suzhou (+2.3m) in Jiangsu province, Xiamen (+1.6m) in Fujian province, and Qingdao (+1.4m) in Shandong province. Note that the census data only released the population changes in each city and do not document detailed migration information. To add some supporting evidence, we further verified the spatial hierarchical migration pattern using a granular mobile phone dataset (fig. S3).

**Mega-cities with massive in-migration are sprawling, giving rise to integrated metropolitan areas.** Population influx has spread from mega-cities to the surrounding counties and cities, e.g., the Shanghai-Hangzhou metropolitan area and the Guangzhou-Shenzhen metropolitan area (Fig. 1C and D; see fig. S4 for more cases). On one hand, the emergence of metropolitan areas arises from the forces of the 'invisible hand' in urban markets (e.g., agglomeration economies). On the other hand, this can be caused by urban policies, such as China's recent push for the integration of urban clusters (9) and the rapid development of transportation infrastructure, especially high-speed railroads. A well-connected transportation network has expanded China's metropolitan areas, making them larger than those in the U.S. For instance, in the Greater Bay Area (the Guangzhou-Shenzhen area and its seven surrounding cities), the total population reaches 78 million, nearly four times the population of the New York metropolitan area – the most populous metropolitan area in the U.S. Remarkably, we also observed a trend of suburbanization within mega-cities, i.e., the central city population moved to the suburbs (Fig. 1B). This suburbanization pattern may be partly attributed to population control policies in these mega-cities which hope to alleviate the housing affordability crisis, air quality degradation, and urban congestion in central cities. Market-oriented development in the suburbs (e.g., rising private car ownership, decentralizing industries, affordable housing, shopping malls, retail parks, etc.) may also fuel such trends (*10*).

**POPULATION AGING, GENDER IMBALANCE, MIGRATION**

The massive migration documented above is also related to more detailed demographic structural transitions.

For example, with a higher proportion of the aging population than ever before, the standard of living in the industrialized world is severely undermined by a declining working-age population, rising health care costs, and increasing dependency ratio. China, the world's largest developing economy, has also begun to see a rapidly growing aging population. According to census data, the population aged 65 and above (hereafter the elderly population) amounted to 190.6 million (13.5% of the total population) in 2020, versus 118.8 million in 2010 (8.87%). Moreover, the proportion of the aging population varies widely across geography (Fig. 2, fig. S5-S9, and Table S3). For example, the ten most aged cities (counties) have an average of 21.04% (27.18%) of the population over 65 years old, while this share is only 4.66% (3.09%) in the ten "youngest" cities (counties).

Contrary to the common belief that large cities are more aged, we find a weak correlation between city size and population aging (fig. S10). For example, we observe both 'young' big cities like Shenzhen (3.22% 65+) and aging big cities like Shanghai (16.3% 65+) in our data. However, we find that the proportion of the elderly population is strongly correlated with population growth (Fig. 2B). Growing cities have a lower rate of aging than shrinking ones. One possible explanation for this strong correlation is that China's out-migration population is primarily composed of working-age adults and college students, resulting in fewer newborns and an increasing share of elderly people in regions with a net population outflow. For these aging regions, the growing elderly population and the decline in the working-age population (fig. S7) pose a serious challenge for instituting sustainable social insurance programs, pension systems, and retirement funds to ensure retirement income security.

Besides concerns about the aging population, gender imbalance is also a long-standing problem in China. Overall, China's sex ratio has remained virtually unchanged over the past decade, with a male to female ratio standing at 1.05. However, the gender imbalance has worsened geographically. The Gini coefficients of county-level population increase from 0.386 (female) and 0.3833 (male) in 2010 to 0.417 (female) and 0.420 (male) in 2020. The unaligned spatial distribution of male and female populations reduces marriage opportunities, further reducing the country's fertility rate which is already on the decline. Interestingly, we find the city-level gender imbalance is again strongly correlated with the city population growth, rather than population size, i.e., higher population growth is associated with a greater gender imbalance (Fig. 2C and fig. S13). This pattern also appears consistent with the hypothesis that the mobile population is largely dominated by migrant workers – a population mostly composed of men.

Taken together, our analysis reveals some new relationships between migration patterns and population spatial structures. Since many complex factors and processes are intertwined with demographic dynamics, our data do not allow us to conduct a direct causal test of the underlying mechanisms. Yet, these novel patterns at the micro-spatial scale represent a promising direction for a better understanding of the evolving population geography in China. We leave more thorough causal analyses of other underlying mechanisms to future research.

**DISCUSSION AND IMPLICATIONS**

In response to potential demographic crises, China has made several major population policy shifts in the last decade. The country adopted different versions of the two-child policy starting in 2009 and then rapidly shifted to a three-child policy due to the weak policy response among young Chinese. However, many still believe that the aggressive new population policy has a dire prospect of obtaining responses among the younger generation unless the excessive costs of living and raising children are effectively reduced. To this end, the government has made many other efforts to combat the rising cost of living, e.g., the recent harsh crackdown on property market speculation and after-school K-12 education. The impacts of such policies remain to be studied. What lies ahead for China's policy remedies for troubling demographic trends? Our probe into the population census and land data can provide insights into this question.

A key parameter here is the land and housing supply. It is well documented that there is a mismatch between population and land (*11, 12*). To shed light on this misallocation issue, we aggregate land transaction data (SM) for two groups of cities (growing and shrinking). Fig. S14 shows that cities with population inflows received a much higher share of land supply before 2010 than shrinking cities. The initial ample land supply provided a positive shock to total factor productivity (TFP) and economic growth in these growing cities (*13*). The gap in the share of land supply between growing and shrinking cities had narrowed from 2000 to 2005 but has leveled off since 2006. Relatively speaking, growing cities did not receive more construction land quotas and faced tight land supply policies over time despite their increased demand as a result of massive population growth. Consequently, the spatial misallocation of land caused by place-based land policies pushed up land and housing prices, dramatically raising the cost of living in growing cities. We argue that the country as a whole would gain more in terms of population and TFP if it is easy to build more housing and provide urban land in more productive cities to alleviate the affordable housing crisis. Correcting misallocation of land conversion quotas across cities is central to addressing the potential demographic crises.

Discussions of many other policy remedies are also in progress. For instance, affordable housing provision should complement the private commercial housing market to address the housing affordability crisis in growing cities. Also, the vast regional disparity and inequality for the aging population shown in our article suggest substantial fiscal risks and challenges when it comes to financial sustainability and retirement income security. Potential future reforms proposed in the scholarly field, including diverse pension financing vehicles, delayed retirement age, inter-regional transfer payments, and housing wealth for retirement, require more discussions and further research (*14*). Moreover, if the demographic turning point is inevitable as it is in many parts of the world (*15*), the government can reduce the socioeconomic impacts of declining birth rates via promoting rapid human capital accumulation through domestic and global talent policies as well as metropolitan area integration.

China's demographic changes and population geography have important global economic and geopolitical implications. A deeper investigation of the internal migration pattern will offer a unique perspective into understanding how China's demographic changes influence the long-run economic prospect of the world's largest transitional, developing economy.

**Acknowledgements:** We thank Shu Cai, Min Fang, Zibin Huang, and Yian Yin for helpful discussions. We are also grateful to Xiaohuan Wu and Yunhan Yang for their superb research assistance.
**Funding:** Y.L. and L.D. acknowledge the research funding from the National Natural Science Foundation of China (Grant Nos.: 41830645, 41801299).
**Author contributions:** L.D., R.D., and Y.L. designed the research. L.D. analyzed the data. All authors contributed to the interpretation of results and manuscript writing.
**Competing interests:** The authors declare no competing interests.
**Data and materials availability:** Data used in this research are available at https://github.com/leiii/census.

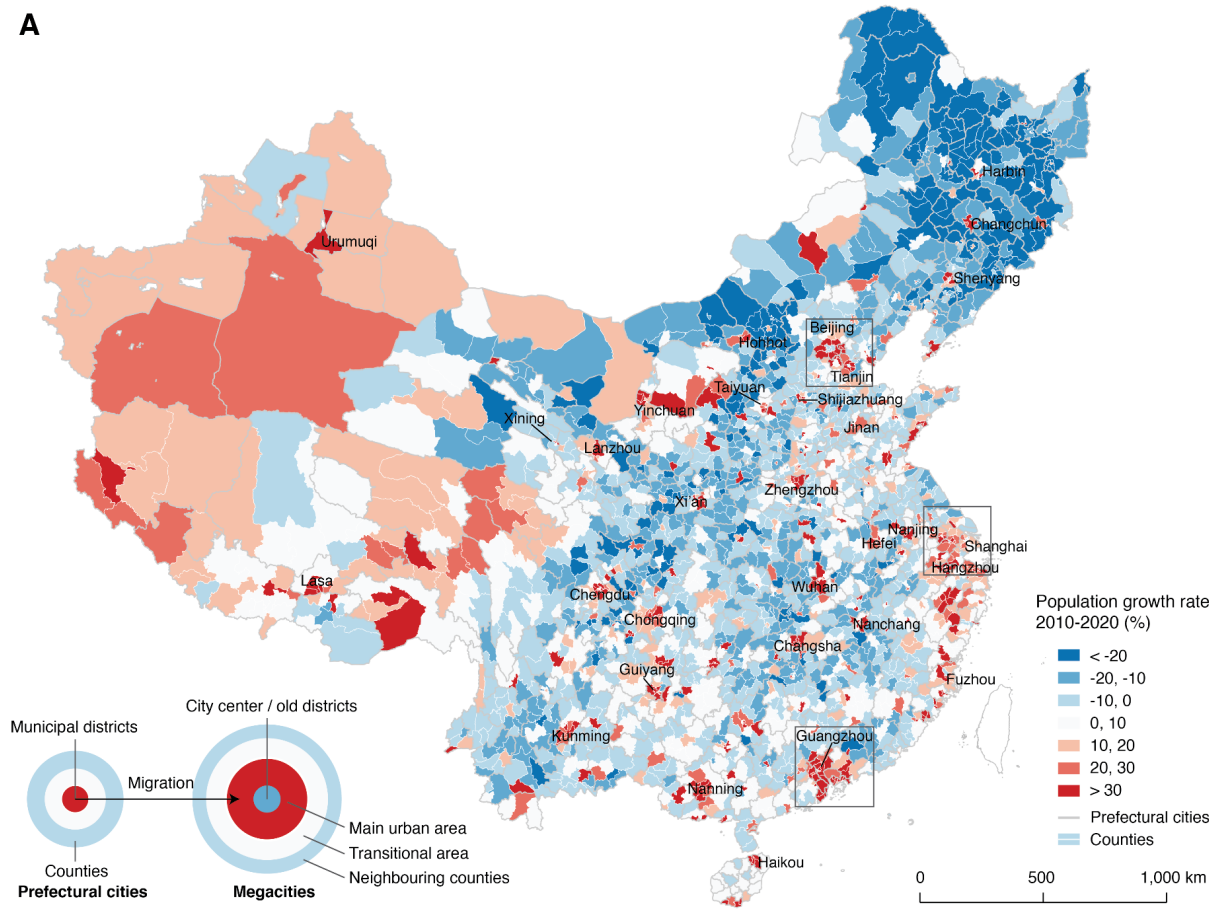
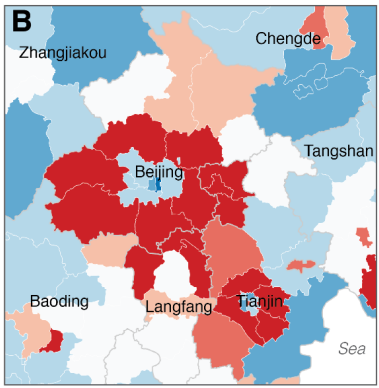
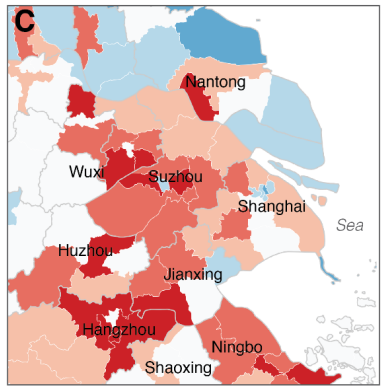
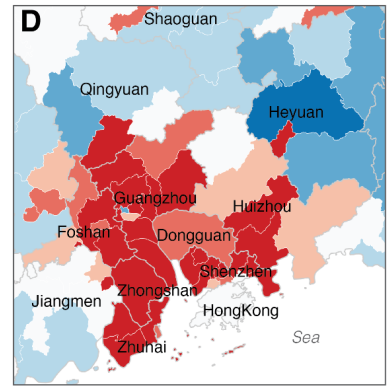

**Fig. 1. Population growth based on data from China's 2010 and 2020 censuses.** (**A**) Nationwide. (**B**) Beijing-Tianjin. (**C**) The Yangtze River Delta. (**D**) The Pearl River Delta. The red and blue colors represent areas of population growth and decline, respectively. The provincial capitals and some megacities can be seen as 'red islands' floating in a blue sea of surrounding counties.

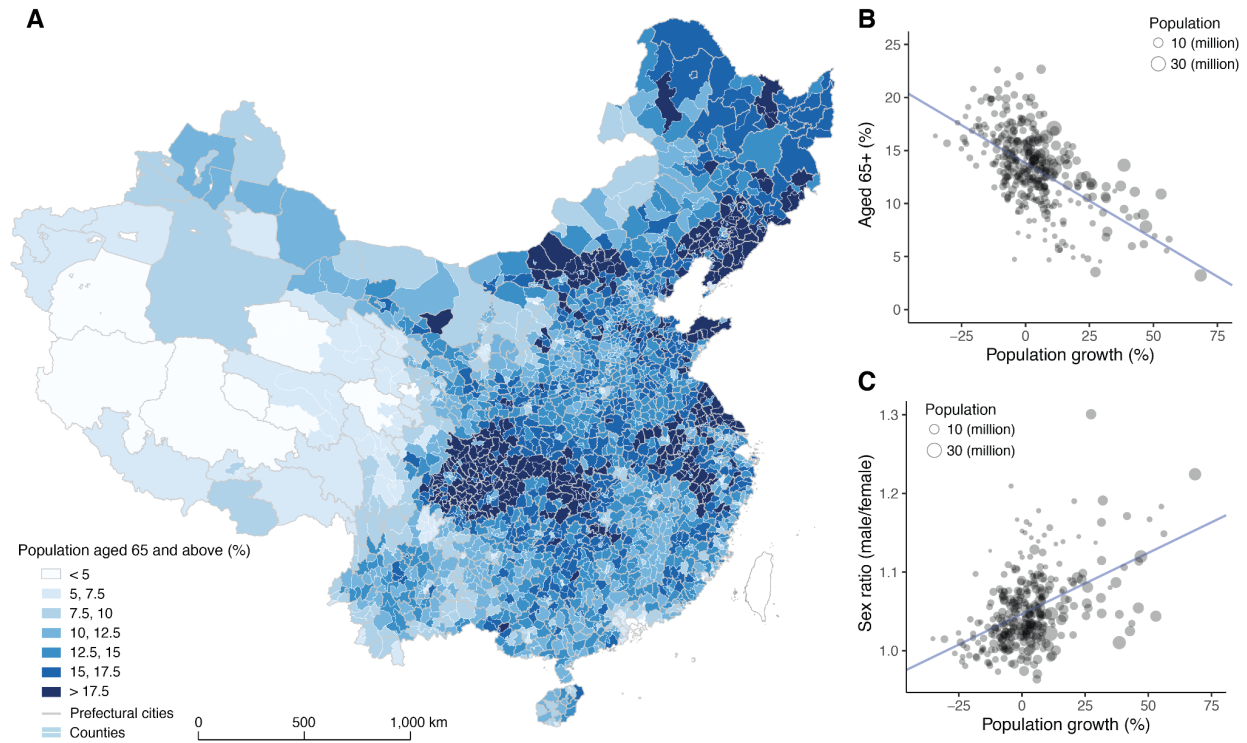

**Fig. 2. The aging rate and gender imbalance based on data from China's 2010 and 2020 censuses**. (**A**) Sichuan and Liaoning provinces face the most serious aging problem due to the substantial outflow of their workforce to more developed regions on the coast. Nine of the top 20 cities with the highest aging rates are in Sichuan and four are in Liaoning. (**B**) A simple linear regression suggests that a one-percentage-point decline in population growth rate would increase the aging rate by 0.14 percentage points at the city level ($R^2 = 0.34$) or 0.09 percentage points at the county level ($R^2 = 0.28$). (**C**) Higher population growth leads to a greater gender imbalance. A one-percentage-point increase in population growth rate would increase the sex ratio by 0.16 percentage points at the city level ($R^2 = 0.23$).

Supplementary Materials for

# Mapping evolving population geography in China:
## Spatial redistribution, regional disparity, and urban sprawl

Lei Dong, Rui Du, Yu Liu

Corresponding author. Email: leidong@pku.edu.cn (L.D.)  liuyu@urban.pku.edu.cn (Y.L.)

**The PDF file includes:**

    Materials and Methods

    Supplementary Text

    Figs. S1 to S13

    Tables S1 to S3

    References

## Materials and Methods

### S1 Data Description

#### S1.1 Census data

The data sources used in this paper are the statistical bulletins of the censuses published by local governments. We collected relevant information from official government websites, the census website, and the bureau of statistics website. We also included census statistics released by some local governments on their official WeChat Accounts. There are a handful of counties that did not release data from the above sources. For these counties, we contacted the local statistical bureaus by email or phone to obtain the data. Fig. S1 shows a sample web page of a census statistical bulletin.

Because most of the counties' census documents are web pages, PDFs, or even images, we manually converted them into a structured dataset; each data point was cross-validated by at least two research assistants to ensure data quality. We corrected a small number of errors in the original information released by the governments.
In addition to the resident population, we also collected data on sex ratio, average number of family members, and age structure (including four groups: 0-14, 15-59, >=60, and >=65) from the statistical bulletins.

Note that there are recently some great efforts made by other scholars to compile the 2020 county-level census data (*1, 2*). However, previous data only include population size and are not open-source. In comparison, our unique contributions are: 1) providing data on a rich set of variables (e.g., population, sex ratio, age groups, family size, etc.); 2) open-sourcing the full dataset; and 3) adjusting the administrative boundary changes based on more accurate first-hand information we gathered (see Section S2).

#### S1.2 Mobile phone data

The mobile phone data (from 2015 to 2016) are obtained from one large online map service provider in China. The location data (longitude, latitude, and timestamps) are generated when users use location-based services (e.g., maps). We aggregate the location points of each user at the city level and update each user's city of residence every three months to get a city-to-city network of migration. As an example, Fig. S3 plots the migration from 2015 Q3 to 2015 Q4.

For the details about the pre-processing and validation of mobile phone location data, please refer to our previous work (*3*).

### S1.3 Land data

Land transaction data are collected from the China Land Market Website (https://landchina.com/), which is maintained by the Ministry of Natural Resources, PRC.

### S1.4 City/county boundary data

The city/county boundary data are collected from https://amap.com, one of the main online map service providers in China.

## S2. Methods

### S2.1 Administrative boundary adjustment

Step 1. For the 2020 Census data, most statistical bulletins show the year-on-year change in population compared to the 2010 Census, which are used to derive the population in 2010 as our baseline population records. In some counties and districts, the bulletins also indicate the administrative boundaries that have been adjusted and provide information about the adjusted population data in 2010 based on the boundary information from the 2020 Census.

Step 2. We compare the 2010 population data from the 2020 Census with those from the 2010 Census. If the records from these two sources are consistent, there is no change in the administrative boundary of the county. Otherwise, the administrative boundary of the county has been adjusted between 2010 and 2020. Note that some districts/counties did not publish the 2010 population data in the 2020 Census, or clearly stated that the boundaries have been adjusted and the 2010 data are unadjusted. We adjust these districts/counties in Step 3.

Step 3. For districts/counties with adjusted administrative boundaries, we further check the 2010 and 2020 statistical codes using the records published on the website of the Bureau of Statistics (http://www.stats.gov.cn/tjsj/tjbz/tjyqhdmhcxhfdm/). We then combine the codes and information from Baidu Baike (China's equivalent of Wikipedia) and government websites (http://www.mca.gov.cn/article/sj/xzqh/1980/ ) to infer the exact adjustment procedure. Common adjustments include changes in administration status (e.g., from county-level city to district), changes in administrative areas, etc.

Step 4. For specific boundary changes, we adjust the census data accordingly. For example, if county A is abolished to establish district A (no change of the administrative area), only the statistical county code is adjusted. If county A and county B are merged into county C, we then add up the population data of county A and county B in 2010 (merged proportional data like population by age structure are the weighted sums of the focal variable in the pre-merging counties). If county D is divided into counties E and F, we then add the population data of counties E and F in 2020. These adjustments require the use of the township-level census data in certain cases. Finally, there are changes present in intercensal years for which adjustment could not be applied due to a lack of information. For example, a portion of county G might be transferred to county H but the transferred area is not well documented in census data. In this case, we merge counties G and H into one county-level unit. The bottom line in the adjustment process is to make the data in 2010 and 2020 comparable.

The statistical summary of the final data is shown in Table S1 and S2.

### S2.2 Gini coefficient

We calculate the Gini coefficient using the equation as follows (*4*):

$G = \frac{\sum_{i=1}^{n}\sum_{j}^{n}|x_i - x_j|}{2n^2(\bar{x})}$, where $x_i$ is the population of city $i$.

## Supplementary Text

In addition to the results in the main text, we also observe another interesting trend in population growth: Not everyone moved to the coasts and the population of western regions has grown significantly. Coastal cities in China are usually magnets for migrants due to better local fundamentals, amenities, job opportunities, trade advantages, and favorable policies. In contrast to the historical trend prior to 2010, most autonomous regions in Western China experienced significant population growth from 2010 to 2020. Tibet experienced the highest growth in population, registering a 17.7% increase over the decade. Xinjiang and Ningxia are also among the top 10 provinces that witnessed massive population growth, registering a growth rate of 14.5% and 12.5%, respectively. This pattern is also present at the city and county levels. For instance, 85.1% of the counties in Tibet and 92.9% of the cities in Xinjiang had a net population gain. In faster-growing western regions, this could be attributed to factors such as China's Western Development Program, reduced travel and trade costs due to rapid transportation infrastructure expansion, poorly enforced one-child policy, and the catch-up effect (i.e., less developed regions grow faster than more developed ones).

# Supplementary Figures

静安区第七次全国人口普查主要数据公报[1]

静安区统计局

静安区第七次全国人口普查领导小组办公室

2021 年 6 月 1 日

根据国务院的决定,我国以 2020 年 11 月 1 日零时为标准时点进行了第七次全国人口普查。在市政府、区委区政府的统一领导和全区市民的配合下,通过全区 5000 多名普查工作人员的艰苦努力,圆满完成普查现场登记和复查任务。现将普查主要数据公布如下:

一、全区常住人口

全区常住人口[2]为 975707 人,同第六次全国人口普查的 1077284 人相比,十年共减少 101577 人,降低 9.4%。平均每年减少 10158 人,年平均增长率为-1%。

全区常住人口中,外省市来沪常住人口为 256445 人,占比 26.3%,同第六次全国人口普查的 257214 相比,十年共减少 769 人,降低 0.3%。

二、户别人口

全区常住人口中,共有家庭户[3] 376054 户,集体户 29821 户,家庭户人口为 879524 人,集体户人口为 96183 人。平均每个家庭户的人口为 2.34 人,比 2010 年第六次全国人口普查的 2.55 人减少 0.21 人。

三、性别构成

全区常住人口中,男性人口为 475832 人,占 48.8%;女性人口为 499875 人,占 51.2%。常住人口性别比(以女性为 100,男性对女性的比例)由 2010 年第六次全国人口普查的 99.77 降至 95.19。

四、年龄构成

全区常住人口中,0-14 岁人口为 90850 人,占 9.3%;15-59 岁人口为 576866 人,占 59.1%;60 岁及以上人口为 307991 人,占 31.6%,其中 65 岁及以上人口为 214376 人,占 22.0%。

与 2010 年第六次全国人口普查相比,0-14 岁人口的比重提高 2.3 个百分点,15-59 岁人口的比重下降 14.1 个百分点,60 岁及以上人口的比重提高 11.8 个百分点,65 岁及以上人口的比重提高 8.4 个百分点。

五、各种受教育程度人口

全区常住人口中,拥有大学(指大专及以上)文化程度的人口为 395559 人;拥有高中(含中专)文化程度的人口为 234150 人;拥有初中文化程度的人口为 223193 人;拥有小学文化程度的人口为 76815 人(以上各种受教育程度的人包括各类学校的毕业生、肄业生和在校生)。

与 2010 年第六次全国人口普查相比,每 10 万人中拥有大学文化程度的由 28289 人上升至 40541 人;拥有高中文化程度的由 28208 人下降至 23998 人;拥有初中文化程度的由 29880 人下降至 22875 人;拥有小学文化程度的由 8791 人下降至 7873 人。

**Fig. S1. The original document of the census bulletin (Shanghai/Jing'an district).**

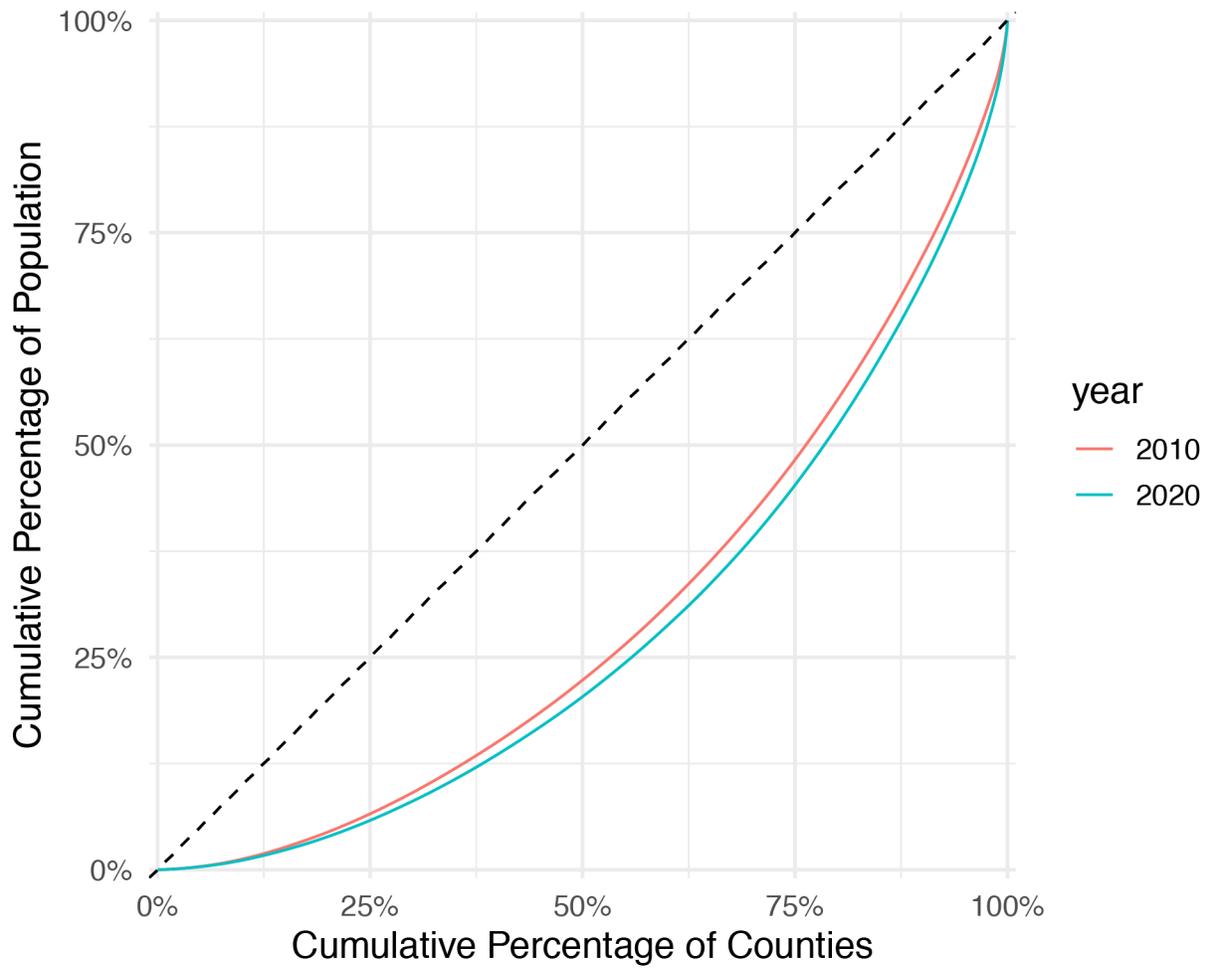

**Fig. S2. Lorentz curves of the population distribution at the county level.** Population in 2020 is more concentrated.

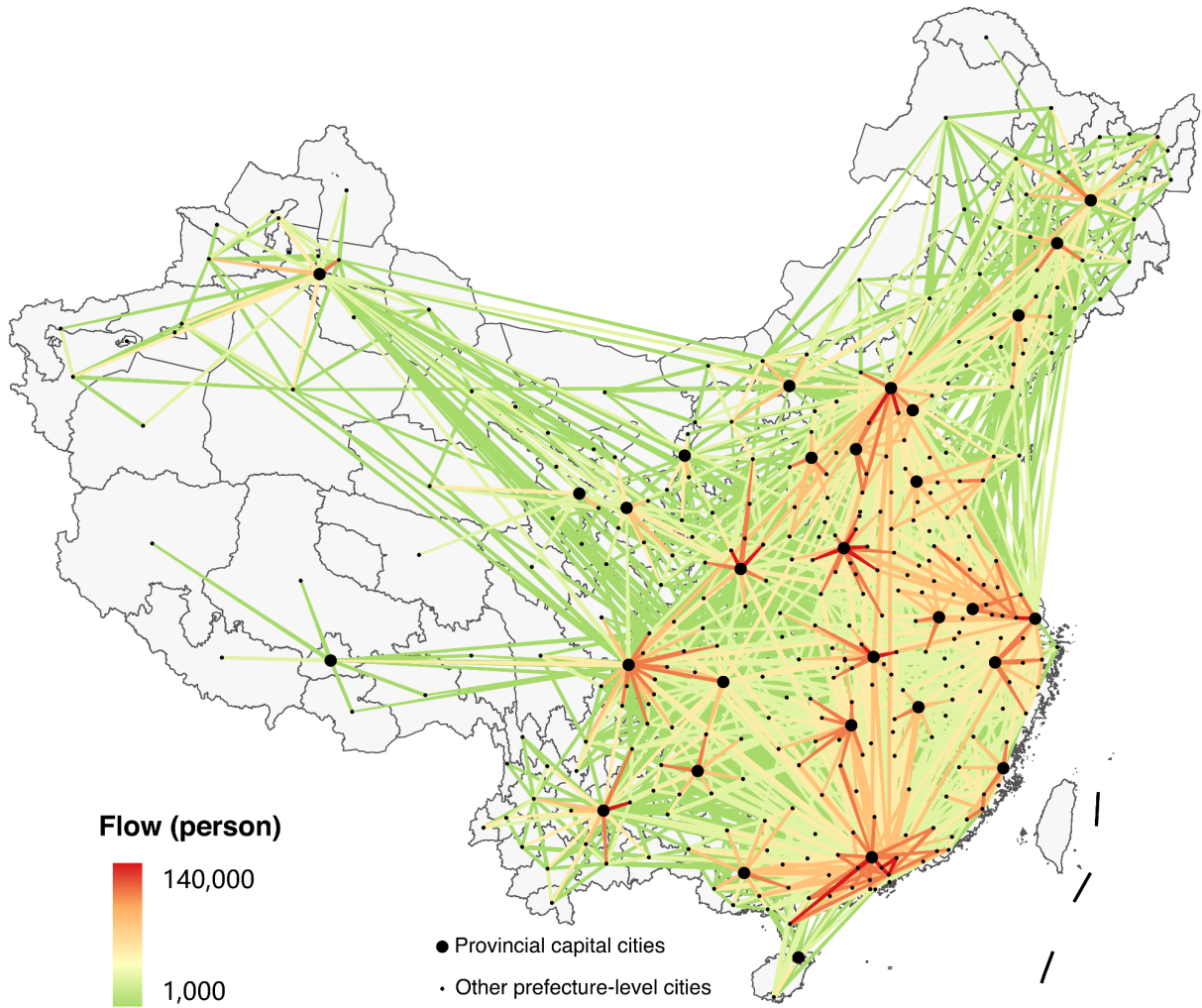

**Fig. S3. Migration detected by mobile phone location data.** The green color indicates the minimal migration flow in our sample, while the red color represents the maximum migration flow. Provincial capitals and municipalities directly under the central government (Beijing, Shanghai, Tianjin, and Chongqing) are symbolized by large black dots, while other prefectural cities are symbolized by smaller black dots.

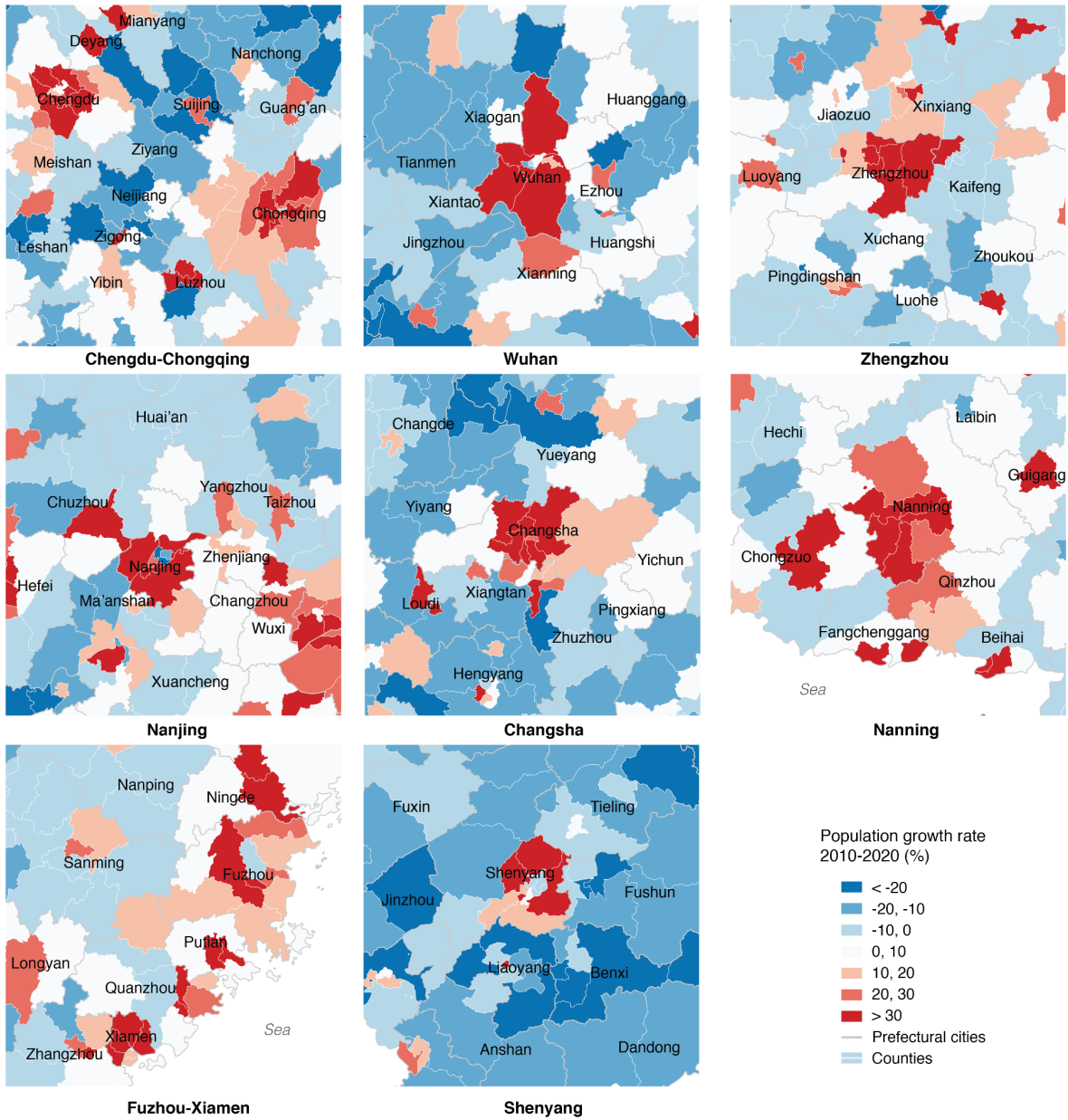

**Fig. S4. County-level population growth of main metropolitan areas based on data from 2010 and 2020 censuses.** The red and blue colors represent areas of population growth and decline, respectively.

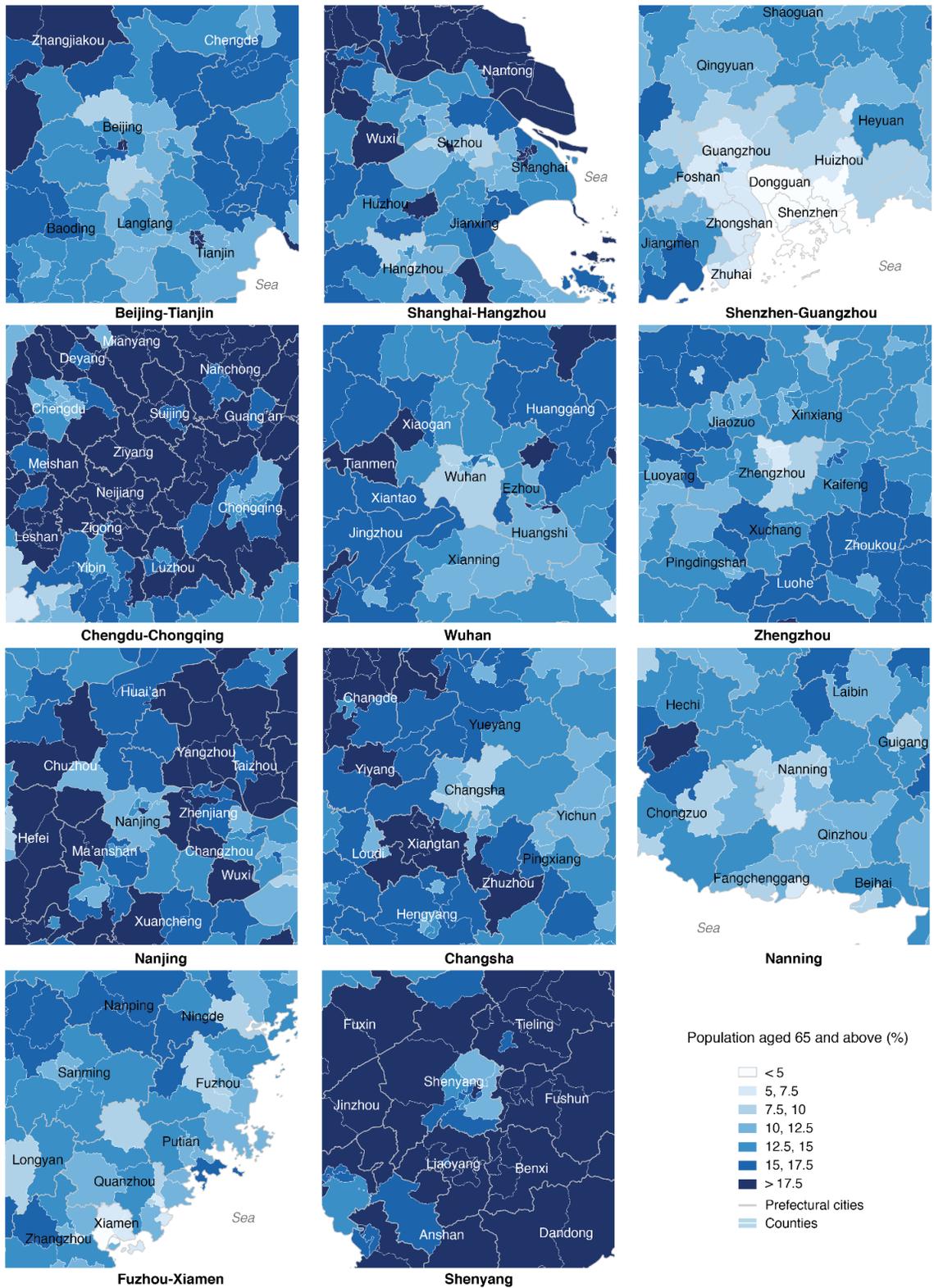

**Fig. S5. County-level aging population (65 years old and above) of main metropolitan areas based on data from the 2020 census.**

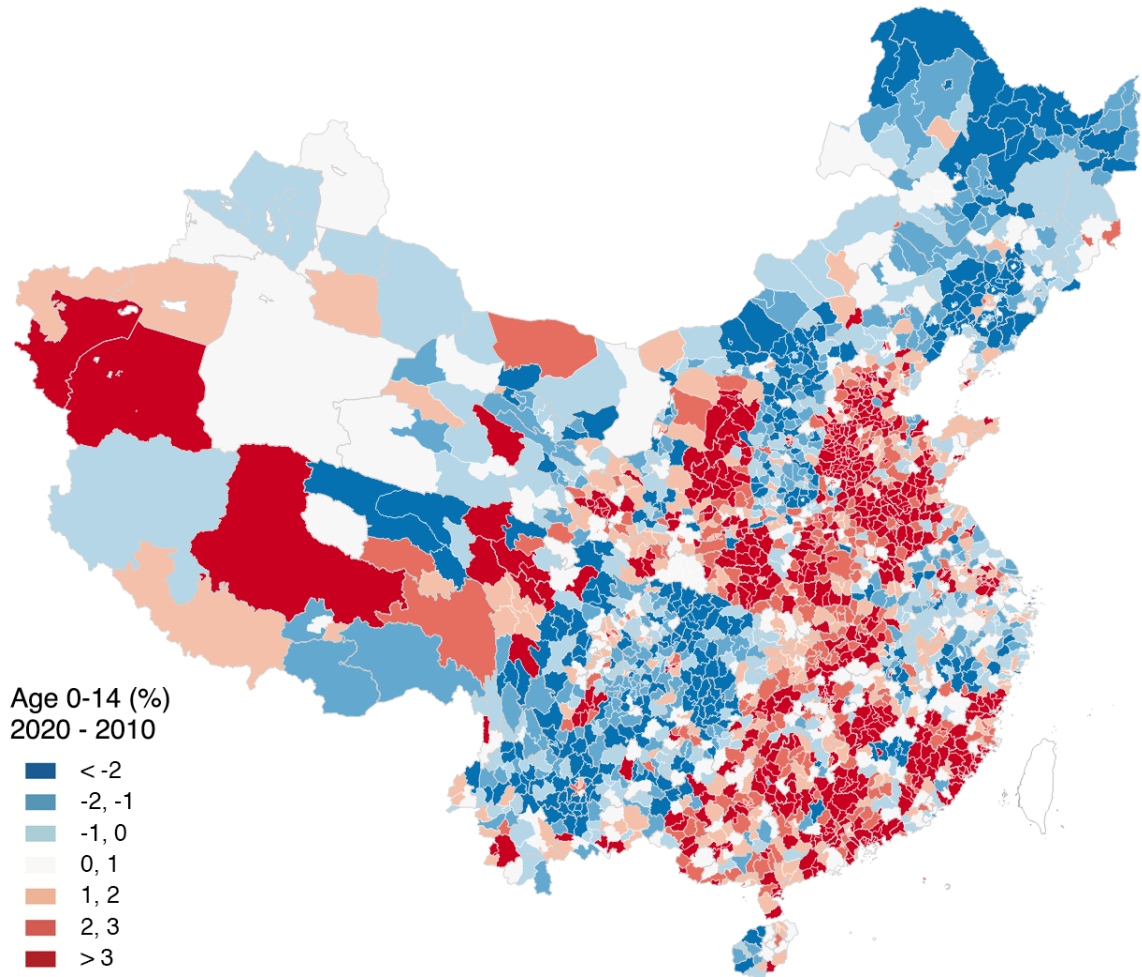

**Fig. S6. The difference in the share of the population aged 0-14 years from 2010 to 2020.**

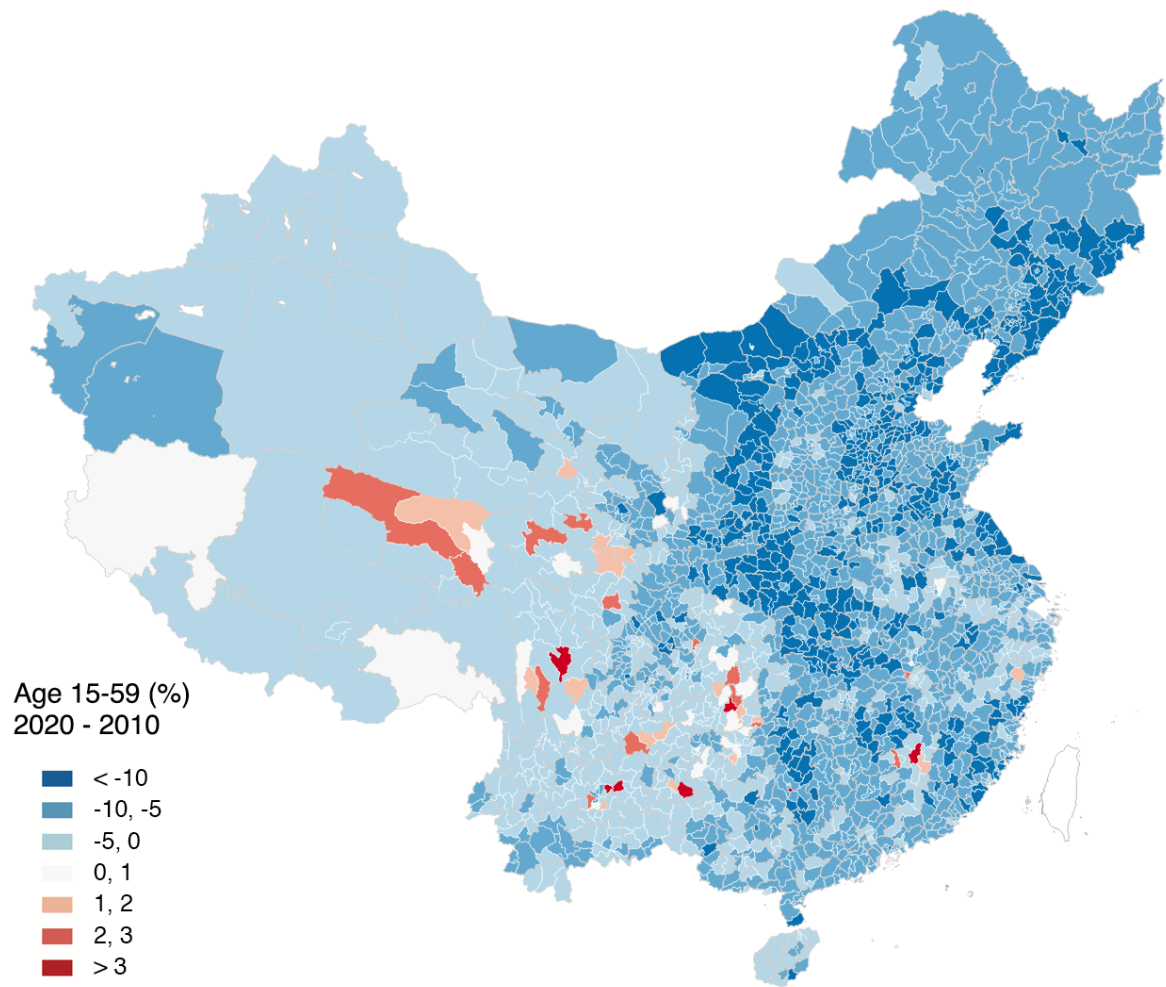

**Fig. S7. The difference in the share of the population aged 15-59 years from 2010 to 2020.** This age group has the most dramatic decline. For better visualization of the data, we use larger intervals for values less than 0.

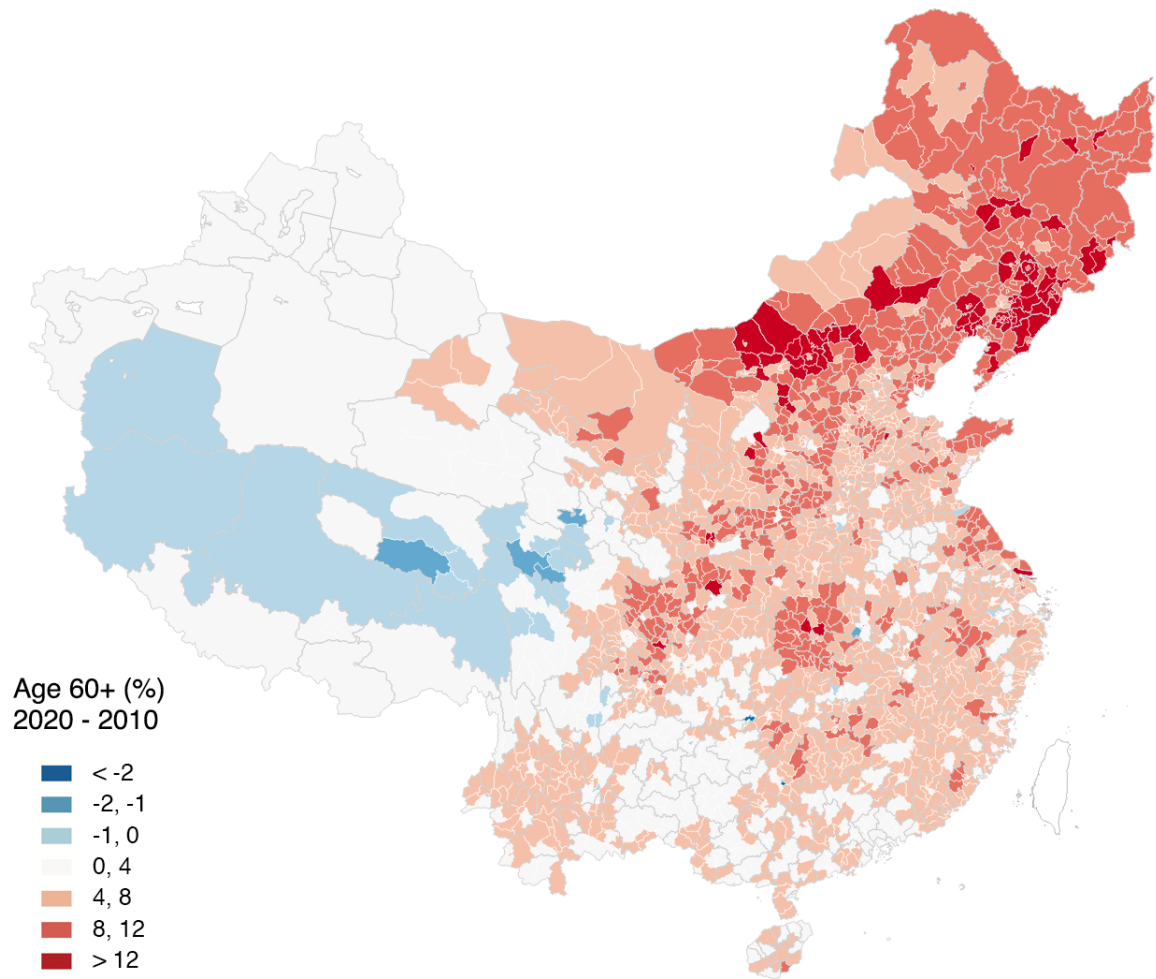

**Fig. S8. The difference in the share of the population aged 60+ years from 2010 to 2020.** For better visualization of the data, we use larger intervals for values greater than 0. The aging problem is most severe in the Northeast, which also has the largest population outflow.

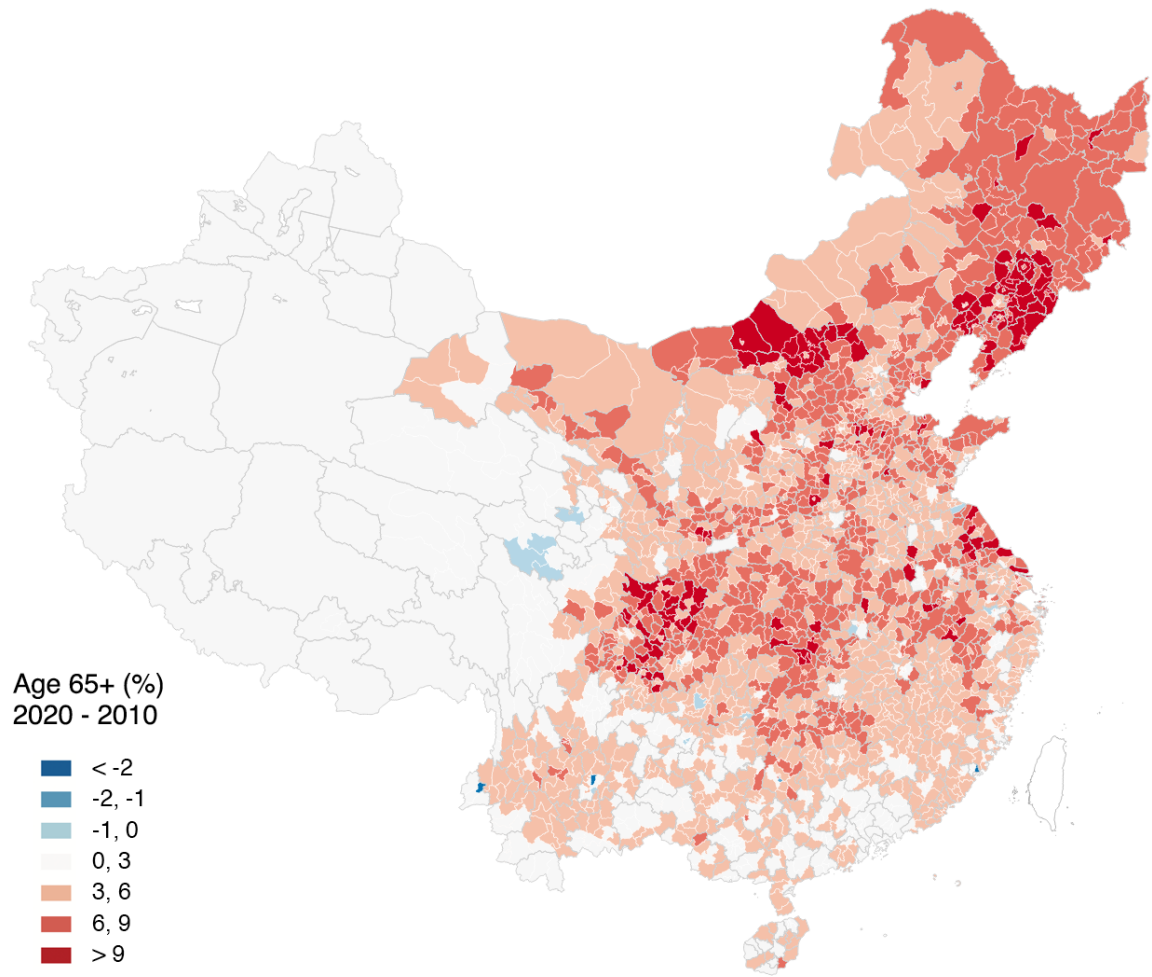

**Fig. S9. The difference in the share of the population aged 65+ years from 2010 to 2020.** For better visualization of the data, we use larger intervals for values greater than 0. The aging problem is most severe in the Northeast, which also has the largest population outflow. The cities with the fastest growth rate of aging population between 2010 and 2020 are mostly located in the three northeastern provinces and Sichuan province, which account for 18 of the top 20 fastest aging cities.

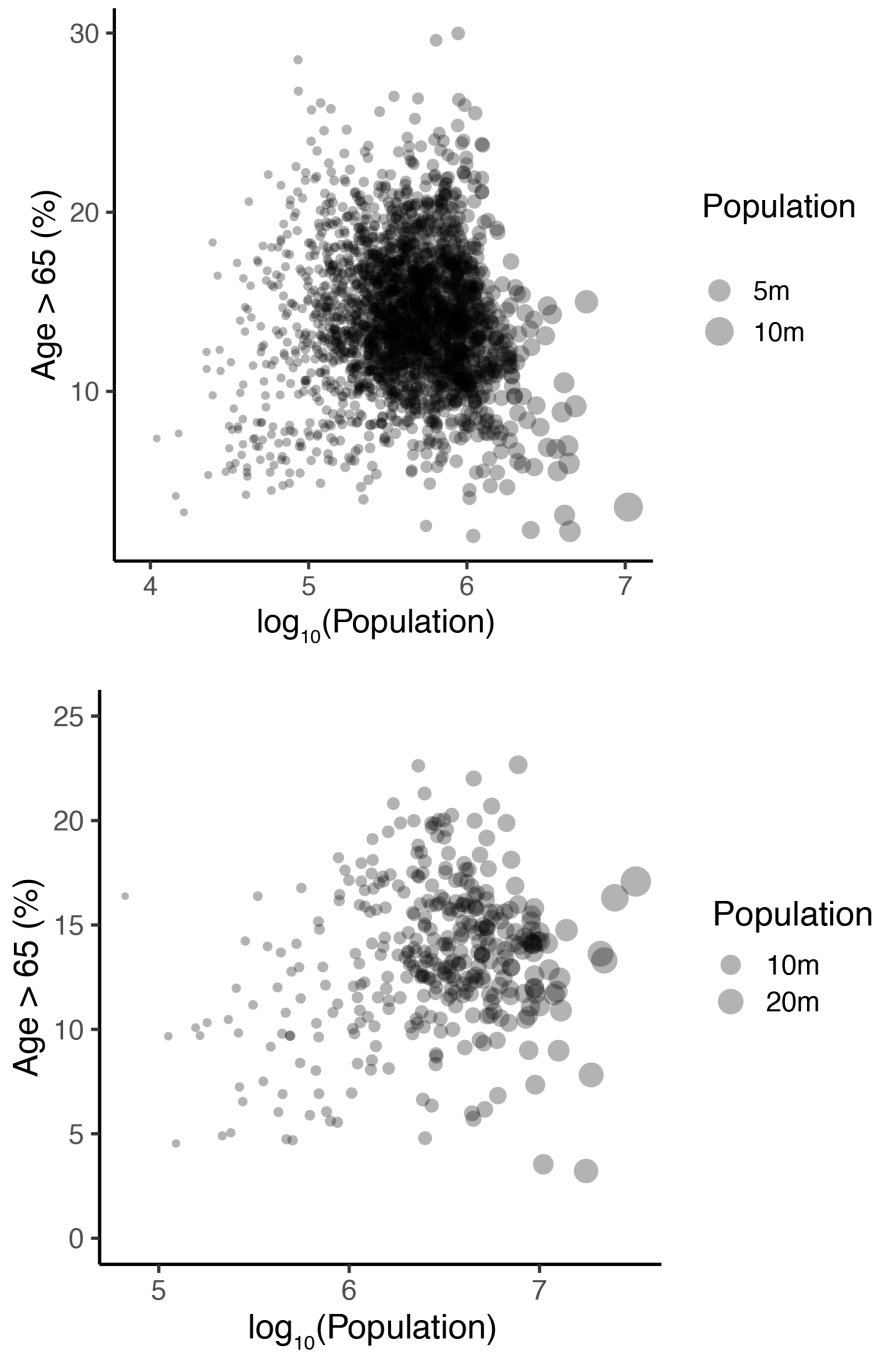

**Fig. S10. City population size and aging population (65+ years old).** The OLS regression shows that the percentage of the aging population is uncorrelated with the county (the upper panel) or city (the lower panel) population size. ($R^2$ = 0.000 at county level and 0.047 at city level, respectively).

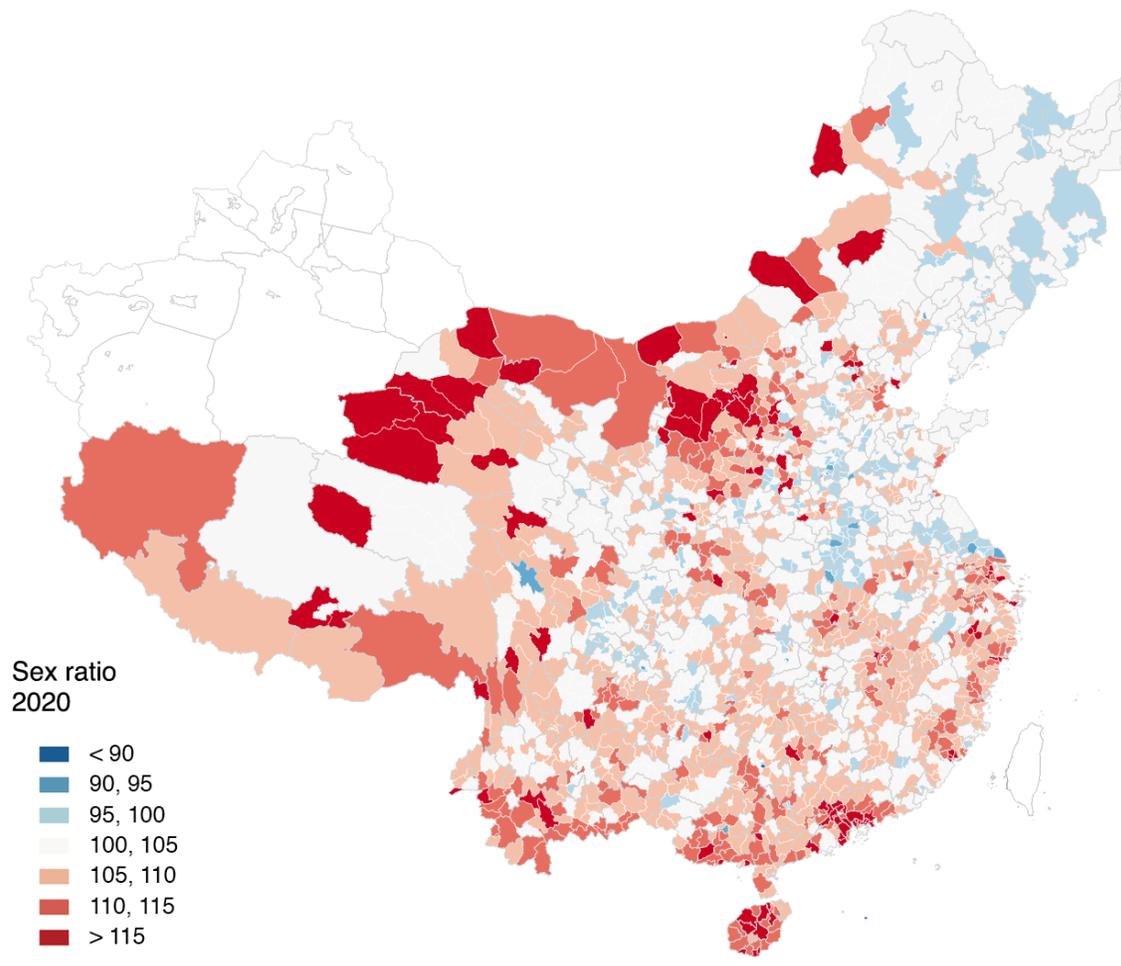

**Fig. S11. Sex ratio in 2020 (male/female; female = 100).**

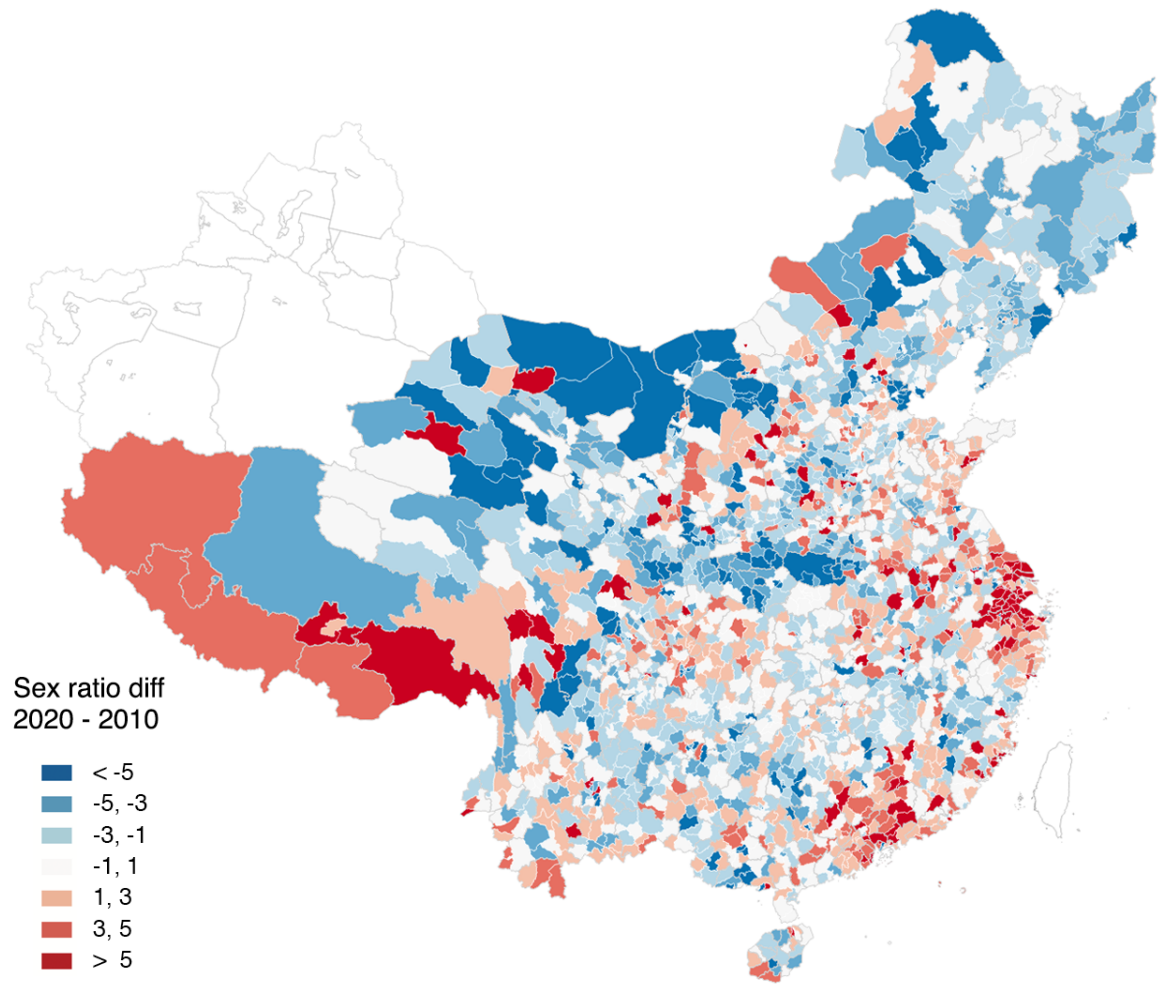

**Fig. S12. The difference in the sex ratio from 2010 to 2020 (male/female; female = 100).**

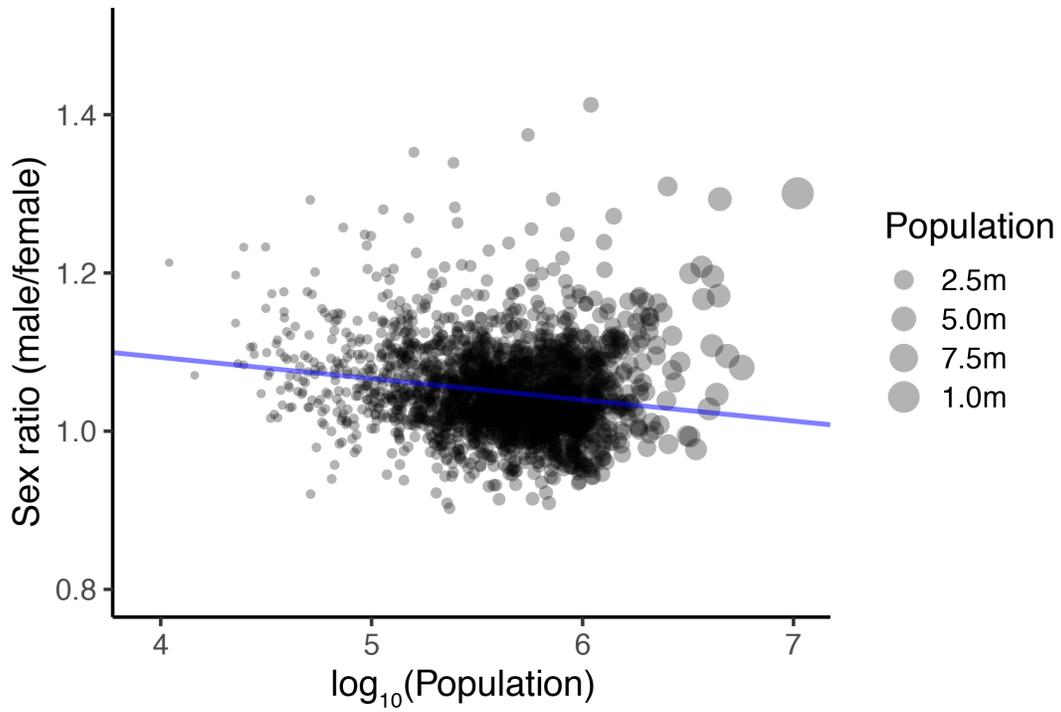
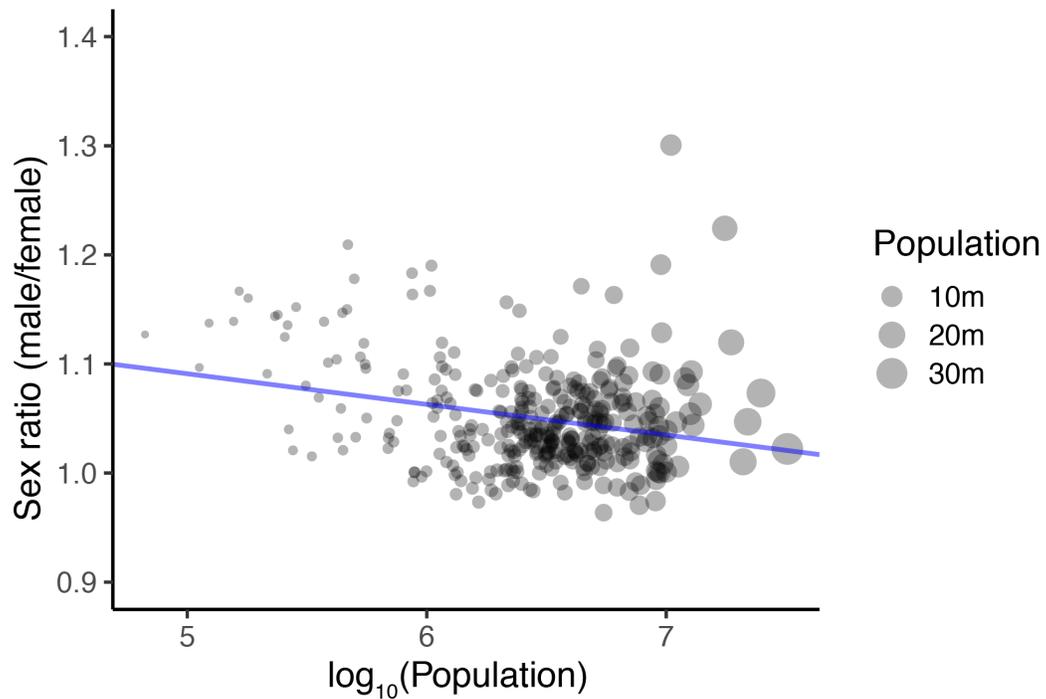

**Fig. S13. Gender imbalance and city size.** $R^2$ = 0.0247 at the county level (upper panel) and $R^2$ = 0.0634 at the city level (lower panel).

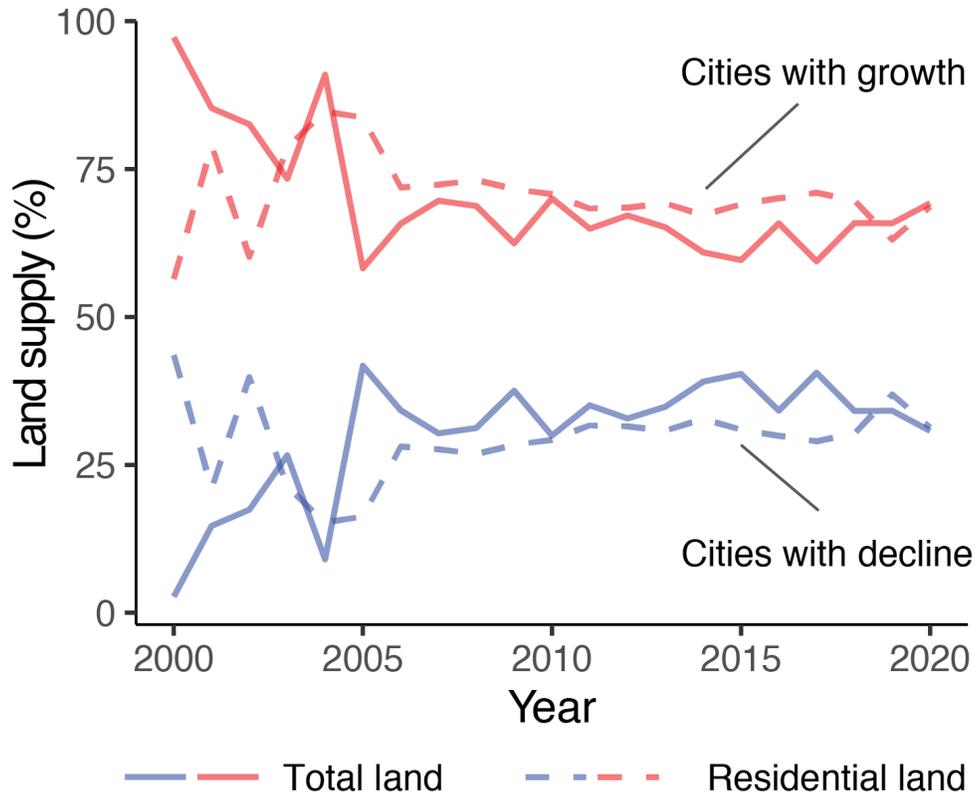

**Fig. S14. Land supply and population growth.** We divide cities into two groups by population change from 2010 to 2020: cities with net population growth (growing cities) and loss (shrinking cities). Red and blue colors represent cities with net population growth and decline, respectively. The two types of line represent total land areas (solid lines) and residential land areas (dashed lines), respectively. Although the population size of these growing cities has been increasing, their land supply as a percentage of the total urban land supply has remained virtually unchanged since 2010.

# Supplementary Tables

**Table S1: City-level descriptive statistics**

| Variables | Obs. | Mean | Median | Std. Dev. | Min. | Max. |
|---|---|---|---|---|---|---|
| Population (2020) | 356 | 3,949,005 | 2,999,886 | 3,747,933 | 66,571 | 32,054,159 |
| Population (2010) | 356 | 3,732,501 | 3,016,123 | 3,207,296 | 76,140 | 28,846,170 |
| Population growth (%) | 356 | 3.14 | 1.69 | 14.67 | -35.24 | 68.46 |
| Household size (2020) | 333 | 2.69 | 2.67 | 0.345 | 1.88 | 4.17 |
| Household size (2010) | 355 | 3.21 | 3.17 | 0.467 | 2.12 | 5.55 |
| Sex ratio (male/female, female = 100, 2020) | 341 | 105.08 | 104.18 | 4.75 | 96.37 | 130.06 |
| Sex ratio (male/female, female = 100, 2010) | 355 | 105.96 | 105.18 | 4.61 | 89.65 | 132.41 |
| Age 0-14 (%, 2020) | 355 | 18.37 | 18.07 | 5.01 | 7.40 | 34.34 |
| Age 0-14 (%, 2010) | 356 | 17.44 | 16.95 | 4.63 | 8.25 | 31.86 |
| Age 15-59 (%, 2020) | 355 | 63.06 | 62.22 | 4.34 | 54.99 | 81.41 |
| Age 15-59 (%, 2010) | 356 | 69.67 | 69.73 | 4.44 | 57.27 | 88.19 |
| Age 60+ (%, 2020) | 355 | 18.58 | 18.41 | 4.77 | 5.36 | 30.26 |
| Age 60+ (%, 2010) | 356 | 12.89 | 12.98 | 2.80 | 3.00 | 23.45 |
| Age 65+ (%, 2020) | 355 | 13.37 | 13.53 | 3.58 | 3.22 | 22.67 |
| Age 65+ (%, 2010) | 356 | 8.69 | 8.70 | 2.10 | 1.83 | 22.70 |

*Note:* The city-level administrative units include 'prefectural cities' (*dijishi and diqu*) and municipalities directly administered by the central government.

**Table S2: County-level descriptive statistics**

| Variables | Obs. | Mean | Median | Std. Dev. | Min. | Max. |
|---|---|---|---|---|---|---|
| Population (2020) | 2,666 | 518,160 | 392,658 | 520,059 | 8,454 | 10,466,625 |
| Population (2010) | 2,666 | 490,568 | 393,331 | 422,945 | 6,883 | 8,220,207 |
| Population growth (%) | 2,666 | 2.58 | -1.61 | 24.43 | -73.93 | 372.96 |
| Household size (2020) | 1,794 | 2.69 | 2.65 | 0.349 | 1.74 | 4.34 |
| Household size (2010) | 2,517 | 3.18 | 3.15 | 0.482 | 1.00 | 5.59 |
| Sex ratio (male/female, female = 100, 2020) | 2,353 | 105.07 | 104.34 | 6.27 | 77.75 | 237.76 |
| Sex ratio (male/female, female = 100, 2010) | 2,522 | 105.58 | 104.84 | 6.36 | 72.77 | 229.58 |
| Age 0-14 (%, 2020) | 2,399 | 18.39 | 18.05 | 5.26 | 5.77 | 40.78 |
| Age 0-14 (%, 2010) | 2,525 | 17.12 | 16.62 | 5.06 | 6.09 | 36.71 |
| Age 15-59 (%, 2020) | 2,399 | 62.11 | 61.54 | 5.06 | 50.54 | 88.60 |
| Age 15-59 (%, 2010) | 2,513 | 69.54 | 69.72 | 5.14 | 52.02 | 91.86 |
| Age 60+ (%, 2020) | 2,398 | 19.50 | 19.19 | 5.10 | 3.43 | 39.87 |
| Age 60+ (%, 2010) | 2,513 | 13.32 | 13.17 | 2.97 | 1.45 | 27.00 |
| Age 65+ (%, 2020) | 2,398 | 14.08 | 13.94 | 3.86 | 1.93 | 29.98 |
| Age 65+ (%, 2010) | 2,524 | 8.93 | 8.76 | 2.17 | 0.78 | 19.00 |

*Note:* The county-level administrative units include 'county' (*xian*), 'county-level city' (*shi*), 'qi', and 'district' (*qu*). The most populous county-level city in both 2020 and 2010 is Dongguan, which is a prefectural city that should govern county-level administrative districts but instead operates as a county-level city that only governs township-level administrative districts. Therefore, we treat Dongguan (and other ones in a similar situation) as a county-level city.

**Table S3: Aging cities in 2020**

| "Aging" city (province) | Aged 65+ (%) | "Young" city (province) | Aged 65+ (%) |
|---|---|---|---|
| Nantong (Jiangsu) | 22.67 | Shenzhen (Guangdong) | 3.22 |
| Ziyang (Sichuan) | 22.62 | Dongguan (Guangdong) | 3.54 |
| Taizhou (Jiangsu) | 22.01 | Ali (Tibet) | 4.54 |
| Zigong (Sichuan) | 21.29 | Nagqu (Tibet) | 4.69 |
| Ulanqab (Inner Mongolia) | 20.81 | Haixi (Qinghai) | 4.74 |
| Nanchong (Sichuan) | 20.69 | Hetian (Xinjiang) | 4.79 |
| Deyang (Sichuan) | 20.25 | Guoluo (Qinghai) | 4.90 |
| Neijing (Sichuan) | 20.03 | Nyingchi (Tibet) | 5.05 |
| Meishan (Sichuan) | 20.02 | Lasa (Tibet) | 5.55 |
| Yangzhou (Jiangsu) | 19.99 | Shigatse (Tibet) | 5.61 |
| Dandong (Liaoning) | 19.99 | Kashgar (Xinjiang) | 5.72 |
| Yancheng (Jiangsu) | 19.88 | Kizilsu Kyrgyz (Xinjiang) | 5.89 |
| Fushun (Liaoning) | 19.88 | Zhongshan (Guangdong) | 5.98 |
| Jinzhou (Liaoning) | 19.87 | Yushu (Qinghai) | 6.04 |
| Suijing (Sichuan) | 19.85 | Chamdo (Tibet) | 6.06 |
| Bazhong (Sichuan) | 19.67 | Xiamen (Fujian) | 6.17 |
| Guangan (Sichuan) | 19.57 | Aksu (Xinjiang) | 6.35 |
| Liaoyang (Liaoning) | 19.46 | Huangnan (Qinghai) | 6.56 |
| Weihai (Shandong) | 19.16 | Zhuhai (Guangdong) | 6.64 |
| Leshan (Sichuan) | 19.19 | Huizhou (Guangdong) | 6.83 |